\documentclass[%
 reprint,
superscriptaddress,
 amsmath,amssymb,
 aps,
prl,
]{revtex4-1}

\usepackage{graphicx}
\usepackage{dcolumn}
\usepackage{bm}

\begin{document}

\preprint{APS/123-QED}

\title{Extreme cavity expansion in soft solids: damage without fracture}


\author{Jin Young Kim}
\affiliation{%
School of Advanced Materials Science and Engineering, SKKU Advanced Institute of Nanotechnology (SAINT), Sungkyunkwan University, Suwon 16419, South Korea}%


\author{Zezhou Liu}
\affiliation{%
Department of Mechanical and Aerospace Engineering, Field of Theoretical and Applied Mechanics, Cornell University, Ithaca, NY 14853, USA}%

\author{Byung Mook Weon}
\affiliation{%
School of Advanced Materials Science and Engineering, SKKU Advanced Institute of Nanotechnology (SAINT), Sungkyunkwan University, Suwon 16419, South Korea}

\author{Tal Cohen}
\affiliation{%
Massachusetts Institute of Technology, Cambridge, MA 02139, USA}%

\author{Chung-Yuen Hui}
\affiliation{%
Department of Mechanical and Aerospace Engineering, Field of Theoretical and Applied Mechanics, Cornell University, Ithaca, NY 14853, USA}%

\author{Eric R. Dufresne}
\author{Robert W. Style}
\email{robert.style@mat.ethz.ch}
\affiliation{
 Department of Materials, ETH Z\"{u}rich, Z\"{u}rich 8093, Switzerland}%


\date{\today}

\begin{abstract}
Cavitation is a common damage mechanism in soft solids.
Here, we study this using a phase-separation technique in stretched, elastic solids to controllably nucleate and grow small cavities by several orders of magnitude.
The ability to make stable cavities of different sizes, as well as the huge range of accessible strains, allows us to systematically study the early stages of cavity expansion.
Cavities grow in a scale-free manner, accompanied by irreversible bond breakage that is distributed around the growing cavity, rather than being localized to a crack tip.
Furthermore, cavities appear to grow at constant driving pressure.
This has strong analogies with the plasticity that occurs surrounding a growing void in ductile metals.
In particular we find that, although elastomers are normally considered as brittle materials, small-scale cavity expansion is more like a ductile process.
Our results have broad implications for understanding and controlling failure in soft
solids.

\end{abstract}

\pacs{Valid PACS appear here}
\maketitle

Cavitation plays a key role in the failure of solids.
This has long been appreciated in ductile metals, where void/cavity nucleation, growth and coalescence govern the initiation of fracture  and fatigue (\emph{e.g.} \cite{rice69,gurs77,raay19,bai19}).
Thus there is an extensive body of literature devoted to the topic (\emph{e.g.} \cite{tayl48,mccl68, tver81, need92a,hang92}).
Cavitation also occurs in highly elastic  materials,  such as rubber \cite{gent59,gent91,cret16}.
In these systems, cavitation underpins processes ranging from fracture and the failure of adhesives \cite{gent59,kend75,hui03,vill15,cret16},   to  traumatic brain injury\cite{pars06,zimb10,haqu12}.
Furthermore, cavitation by the injection of fluid is emerging as a method to characterize soft materials \cite{zimb07,kund09,zimb10,raay19}.

However, understanding soft-solid cavitation has not been a simple question of extending results from the ductile metal literature.
Researchers have typically treated cavitation in soft solids and ductile metals as separate problems, as these materials have very different properties.
Metals are orders of magnitude stiffer than elastomers and gels. Ductile metals yield plastically at low strains, while soft solids can often stretch elastically to many times their original length before irreversible bond breakage occurs. 
Furthermore, metals are ductile, while elastomers are generally considered as being brittle.

Thus, while void growth in metals is well understood, there is still a lack of consensus on the mechanisms governing soft-solid cavitation (\emph{e.g.} \cite{will65,hutc16,poul17,kang17}).
As a singular event in space and time, cavitation pushes theory and experiment to their limits.
Theoretical challenges arise primarily from the enormous deformations around the expanding cavity, which lead, among other difficulties, to a lack of valid, reliable constitutive models.
Experimental challenges revolve around the fact that cavity growth typically occurs unstably (i.e. fast), and at very small scales.
Thus it is hard to achieve sufficient spatial and temporal resolution to resolve cavity inflation and the separation of elastic, inelastic, and viscous contributions.

\begin{figure}
\centering
\includegraphics[width=1\linewidth]{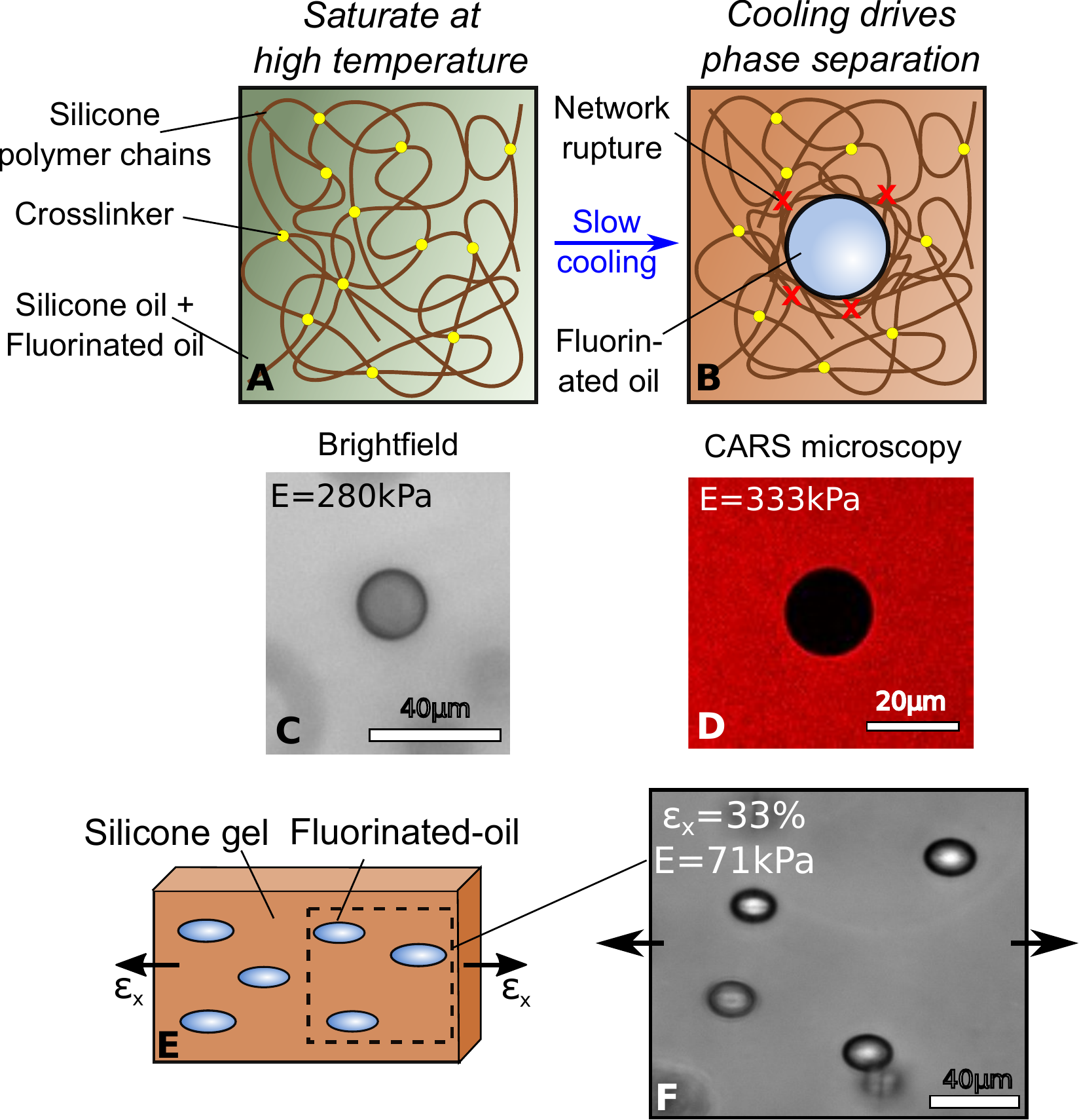}
\caption{Droplet growth via phase separation in silicone gels. A,B) Schematics showing how droplets are formed. Silicone gels are submerged at 40$^\circ$C in fluorinated oil for several hours, until the gels are completely saturated. Upon slow cooling to room temperature, droplets appear. C) A droplet growing in unstretched silicone with $E=$280kPa.
D) A similar droplet in silicone with $E=333$kPa, imaged with CARS microscopy which visualizes the presence of the silicone network. This is clearly excluded from the growing droplet. E,F) When a silicone gel is held with a constant, uniaxial stretch during the phase separation process, droplets grow as spheroids.}
\label{fig:schematic}
\end{figure}

Here, we resolve some of these experimental difficulties by  condensing liquid droplets in soft materials \cite{styl18}.
This approach allows us to slowly and systematically grow and shrink liquid-filled cavities inside unfilled elastomers, without initial defects due to injection.
Breaking symmetry with a macroscopic strain, we are able to easily visualize growth-induced damage. 
We find that small-scale cavity growth has much more in common with ductile metal cavitation than expected.
In particular, at these scales, cavity growth in soft solids is rather like a ductile process, as bond breakage (stress softening) is distributed around the surface of the cavity, instead of being localized to a well-defined crack tip.
This has important implications for understanding and controlling failure in soft solids.

\subsection{Volume-controlled cavity growth}

We nucleated and grew liquid inclusions in silicone gels using the technique shown schematically in Figure \ref{fig:schematic}(a-b) \cite{styl18}.
We created silicone gel samples by mixing silicone polymer chains with different amounts of cross-linker to produce gels with a range of Young's moduli from $E=71-800$kPa.
The resulting gels are highly elastic, showing no evidence of a Mullins effect up to the point of failure in tensile tests consisting of repeated loading/unloading cycles of increasing amplitude (see Supplemental Information).
The gels were immersed in a fluorinated oil (Fluorinert FC770, Fluorochem) that is partially soluble ($\sim 3\mathrm{vol}\%$ at room temperature \cite{roso19}) in silicone, and then incubated at 40$^\circ$C for several hours to allow sample saturation.
Upon slow cooling to room temperature (23$^\circ$C), phase separation occurs, causing nucleation and growth of fluorinated-oil droplets within the silicone gel over a timescale of tens of minutes (e.g. Figure \ref{fig:schematic}c).
By controlling temperature, we effectively have direct control of droplet volume.
Depending on various parameters (chiefly $E$ and the cooling rate), the droplets can grow as large as several tens of micrometers in radius \cite{styl18}.
They are then stable until diffusion of the oil out of the the edges of the sample eventually causes them to shrink and disappear.
Note that $E$ will change slightly during cooling, as described by rubber elasticity theory, but this will be minimal in the range of temperatures that we use.

We confirm that the polymer network is rejected from the droplets with Coherent Anti-Stokes Raman Scattering (CARS) microscopy at a wavenumber of 2912 $\mathrm{cm}^{-1}$ (Figure \ref{fig:schematic}d).
This is a spectroscopic, confocal technique that can detect the vibrational signature of silicone.
The figure shows a typical, fully-grown droplet, displaying the lack of silicone signal inside the droplet.
We find no significant difference between the intensity in such droplets and in pure fluorinated oil, suggesting that the network is fully excluded (see Supplemental Information).

\subsection*{Self-similar, spheroidal droplet growth}

Droplets grown in stress-free gels are always observed to be spherical (see Figure \ref{fig:schematic}c,d and \cite{styl18}).
However, if we pre-stretch the sample with a uniaxial strain $\epsilon_x$, and this stretch is held constant during the entire incubation, nucleation, and growth process, spheroidal droplets form with their long axis parallel to the stretch direction (Figure \ref{fig:schematic}e,f).
As described below, this symmetry breaking gives us information about how damage occurs.

\begin{figure}
\centering
\includegraphics[width=1\linewidth]{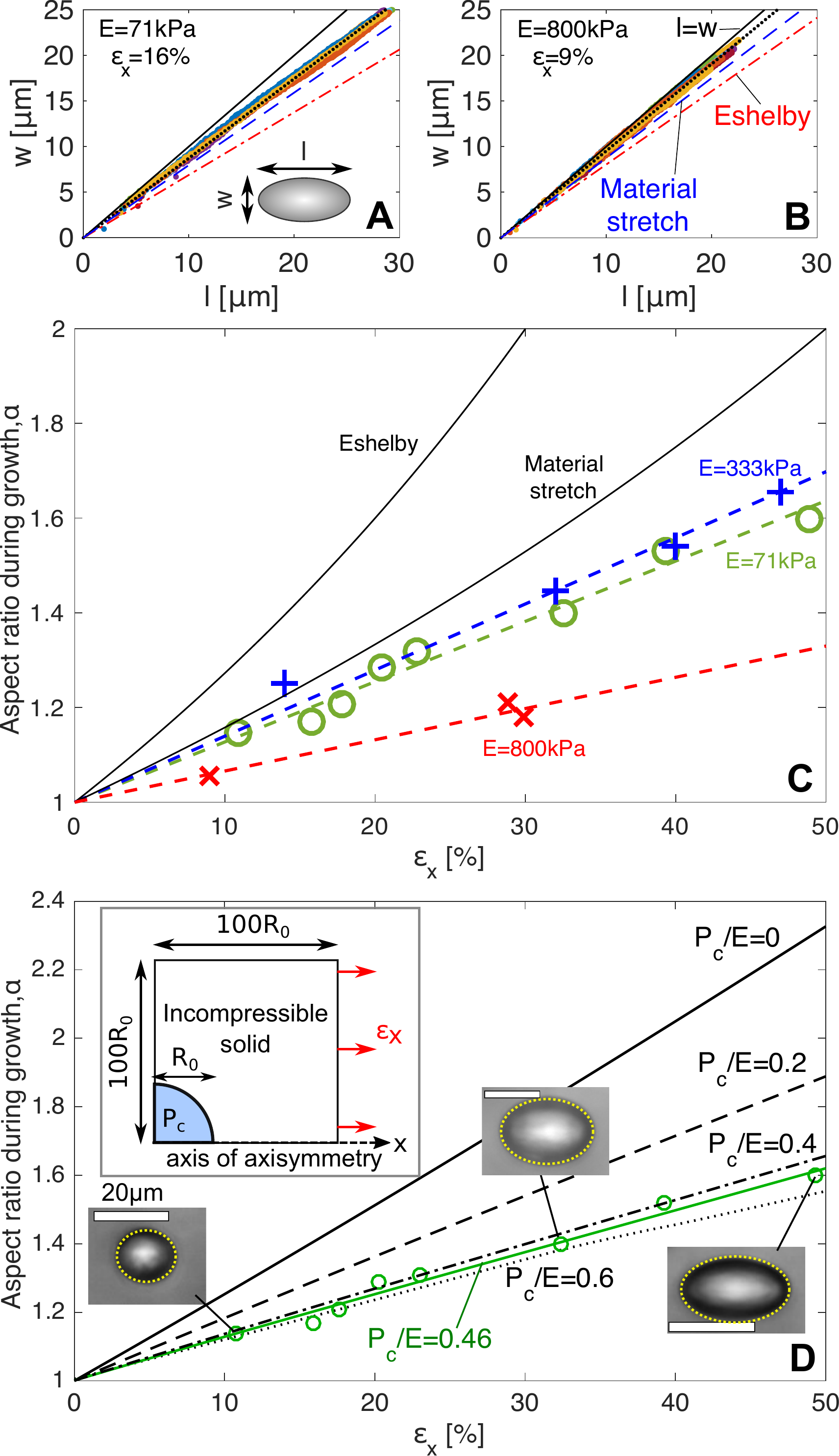}
\caption{Droplet shape evolution during growth. A,B) Droplet growth in different stiffness samples with different applied strains all grow in a scale-free way. Different colors correspond to different droplets. C) The aspect ratio of growing droplets increases linearly with stretch. Dashed lines are lines of best fit, assuming that $\alpha(\epsilon_x=0)=1$ (\cite{styl18}). For all plots we include the predicted shape evolution of droplets from Eshelby theory, and for a volume of solid material that stretches with the surrounding solid.
 D) Results of a theoretical model predicting the relationship between $\alpha$, $P_c$ and $\epsilon_x$.
An inclusion is embedded inside an incompressible neo-Hookean solid with small-strain modulus $E$.
The inclusion is then inflated to, and held at $P_c$ while the solid is stretched with strain $\epsilon_x$.
At fixed $P_c/E$, $\alpha$ essentially is a linear function of $\epsilon_x$.
The green curve shows a fit to the $E=71$kPa data (green circles) from (C).
Images show droplets at different $\epsilon_x$, compared with the model's predictions.
}
\label{fig:selfsimilar}
\end{figure}

For all our experiments, droplets grow with a fixed, spheroidal shape.
Figures \ref{fig:selfsimilar}(a,b) demonstrate the shape evolution of droplets (length, $l$, and width, $w$ during growth) in experiments with different $E$ and $\epsilon_x$.
All of the droplets maintain the same aspect ratio, $\alpha=l/w$ as they grow (see the Supplement for more examples).

We see that there is a strong, linear correlation between $\alpha$ and $\epsilon_x$, with highly elongated droplets forming in the most stretched samples (Figure \ref{fig:selfsimilar}c). 
The aspect ratio also varies with stiffness: droplets growing in the stiffest sample, $E=800$kPa, remain much more spherical than droplets in the two softer samples at the same stretch.
Interestingly, there is a non-monotonic dependence of $\alpha$ on $E$, which suggests that the shape of the droplets is controlled by material parameters beyond $E$ (\emph{i.e.} either non-linear elastic or failure properties).

\subsection{The pressure for droplet growth}

\begin{table}
\centering
\begin{tabular}{ c | c | c | c }
Young's & Fracture & Elasto-adhesive & Elasto-capillary \\
modulus & energy & length & length \\
  $E$ (kPa) & $\Gamma$ (J/$\mathrm{m}^2$) & $\Gamma/E$ ($\mu$m)  & $\Upsilon/E$ (nm)\\
  \hline
  71 & 21 & 300 & 62 \\
  333 & 34 & 102 & 13 \\
  800 & 58 & 73 & 5.5 \\
\end{tabular}
\caption{Measured material properties for the different stiffness silicone gels. $\Upsilon$ is taken as 4.4mN/m, the surface tension of uncured silicone against fluorinated oil.}
\label{table}
\end{table}

We gain useful insight by comparing measured values of $\alpha$ to simple elasticity theory.
For example, Eshelby's inclusion theory \cite{eshe57} predicts that an initially-spherical, incompressible, liquid inclusion, embedded in a linear-elastic solid, will deform as $\alpha=(6+10 \epsilon_x)/(6-5\epsilon_x)$.
However, it dramatically over-predicts measured values (Figure \ref{fig:selfsimilar}).
Indeed, droplets actually appear `stiffer' than the silicone gel:
if we take a uniform piece of material and apply a uniaxial stretch, then its aspect ratio will change to $\alpha=(1+\epsilon_x)/(1-\epsilon_x/2)$.
However the measured value of $\alpha$ is even smaller than this (Figure \ref{fig:selfsimilar}).
One explanation is that there is a significant surface tension, $\Upsilon$, of the droplet interface.
However, we expect this to be negligible, as solid capillarity should only play a role when $w,l\lesssim\Upsilon/E$ \cite{styl15,styl17}.
We estimate $\Upsilon=4.4$mN/m by using the surface tension of uncured polymer chains against the fluorinatedoil, as measured with the pendant droplet method (\emph{e.g.} \cite{dege04}).
This gives a value of $\Upsilon/E$ that is much smaller than all the droplets observed (see Table \ref{table}).

One other explanation why $\alpha$ is not captured by the simple elastic models is that there is a significant pressure, $P_c$, inside the droplet that drives growth, accompanied by large nonlinear deformations.
If this isotropic stress is large in comparison to anisotropic stresses from the macroscopically applied strain, the droplet shape should remain relatively spherical.
We can investigate this effect with a simple model, which reproduces many features of the experiments, even though it does not specifically capture the instability limit.
We treat the growing droplet as an initially-spherical hole in a stretched, non-linear elastic solid, with far-field strain $\epsilon_x$ (Figure \ref{fig:selfsimilar}d).
We inflate the hole with a pressure $P_c$ and measure the resulting shape.
For simplicity, we take the solid to be an incompressible, neo-Hookean material with small-strain elastic modulus $E$.
The droplet (initial radius $R_0$) is embedded in a solid volume 100 times larger than $R_0$. The axisymmetric finite-element model is then solved using ABAQUS software (note that we only need to model a quarter of the domain, due to symmetry).

The results show interesting qualitative agreement with our experiments (Figure \ref{fig:selfsimilar}d, Supplementary Information).
We find that, despite large deformations, if $P_c/E$ is held constant, the computed aspect ratios of the resulting droplet shapes are almost exactly proportional to $\epsilon_x$.
By comparison with Figure \ref{fig:selfsimilar}c, this suggests that droplets in a given material grow with the same internal pressure, independent of the applied strain, $\epsilon_x$.

This suggests that we can try to fit the experimental data for each material in Figure \ref{fig:selfsimilar}c with a unique value of $P_c/E$.
For the $E=71$kPa data, we find $P_c/E=0.46\pm 0.07$.
For the $E=333$kPa data,  $P_c/E=0.32\pm 0.08$.
In these cases, the cavity shapes predicted by the fitted model also agree well with our observations, as demonstrated by the yellow dashed curves superimposed on the images in Figure \ref{fig:selfsimilar}d.
However, for the $E=800$kPa data, it is not possible to fit a value of $P_c/E$.
In this case, the solution becomes unstable as we increase $P_c/E$ (due to the cavitation instability \cite{ball82}) before we find a solution with the correct aspect ratio.
Note that this is not an artefact of using a neo-Hookean material.
We obtain very similar results when we use a 3-parameter Yeoh material that mimics stress-softening (see the Supplement).

The results show that the model captures much of the growth process.
However, there is clearly some missing physics, which means that we cannot use the model to give a quantitative measurement of $P_c/E$.
Instead, we can only assume that the order of magnitude of the fitted values of $P_c/E$ is correct -- so that $P_c\sim E$ during the cavity growth process.
Importantly though, we can conclude that the failure of the elastic model to quantitatively describe the data suggests that there is inelasticity, or damage around droplets, during the growth process.

\subsection{Damage during droplet growth}

\begin{figure}[h]
\centering
\includegraphics[width=0.93\linewidth]{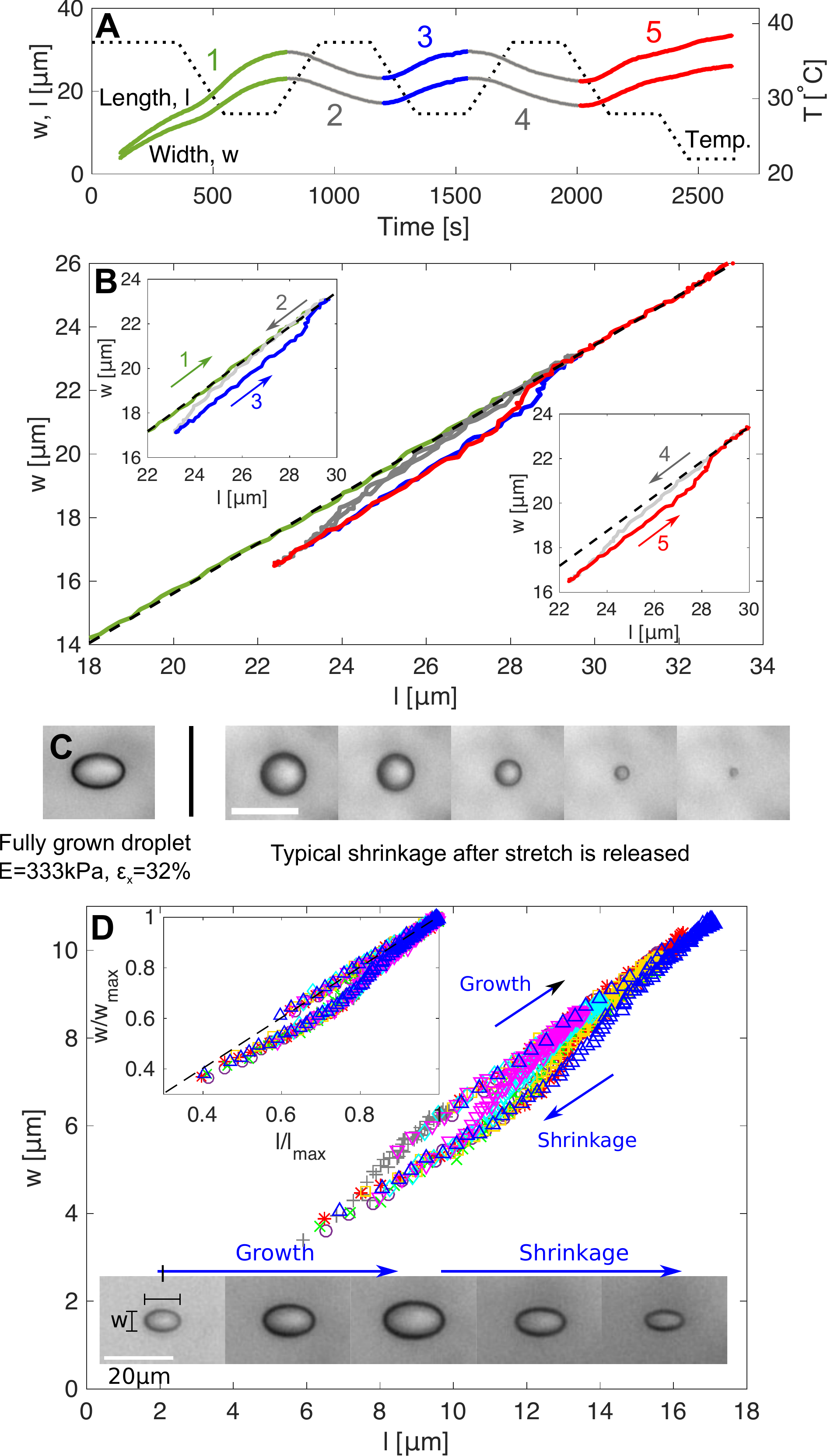}
\caption{Irreversible droplet growth and shrinkage indicates bond breakage.  A) The programmed stage temperature (dotted track), and droplet length and width vs time for a typical droplet in a gel with $E=333$kPa and $\epsilon_x\approx 20\%$. N.B. no droplets nucleate or disappear during the whole cycle. B) The same data plotted with $l$ vs $w$.
The black, dashed line shows constant aspect ratio growth. Insets show the same data, split apart to highlight the cycling behavior.
C) Left: a fully-grown droplet in a stretched silicone. Right: The sample is cut to remove the stretch, and a droplet is imaged as it shrinks.
The scale bar is $20\mu$m.
D) $l$ and $w$ for a selection of droplets as they grow and shrink in silicone with $E=333$kPa and $\epsilon_x=60\%$. Shrinkage is driven by evaporation of fluorinated oil from the sample sides. Different colors correspond to different droplets. Inset: When $l$ and $w$ are re-scaled by the maximum size that a droplet grows to ($l_{\mathrm{max}},w_{\mathrm{max}}$), all tracks collapse onto a single hysteresis curve.
The images show how a typical droplet grows and shrinks.
}
\label{fig:itsfracture}
\end{figure}

We rule out purely elastic growth by examining the irreversibility of droplet growth and shrinkage in a stretched sample.
As a first test, we apply a temperature cycle to a stretched sample ($\epsilon_x\approx 20\%$) in a thermal stage (Instec TSA12Gi).
This causes both $l$ and $w$ to cycle with time, as shown in Figure \ref{fig:itsfracture}a.
Plotting $l$ vs $w$ (Figure \ref{fig:itsfracture}b) then immediately shows evidence of irreversibility: 
during initial growth, droplets grow in a self-similar way.
However, if we then shrink the droplets and regrow them, the shape of the droplet during  regrowth is more elongated (see also images in Figure \ref{fig:itsfracture}d).
If we continue to grow the droplet larger than the size it previously attained, it returns to the same, constant-aspect-ratio growth line that it initially grew along.
This confirms the presence of chain breakage or stress-softening -- even though our silicone gels display no sign of stress-softening in macroscopic tests.
Note that the fact that chain breakage occurs is actually not surprising, as the polymer mesh size of the gel is $O(\mathrm{10nm})$ (\emph{e.g.} \cite{styl18b}), so cavities enlarge by a few orders of magnitude during growth.
This results in extremely large strains that are much bigger than those which can be achieved with simple tensile tests (\emph{e.g.} \cite{lefe15}).

Although droplets grow by a damage mechanism, they differ from brittle fracture in that bond-breakage appears to be distributed around their surface, rather than being localized to a crack tip.
Figure \ref{fig:itsfracture}c shows a typical, fully-grown droplet in a stretched sample.
After droplets have finished growing, the sample is cut to release stress and we observe droplet shrinkage, as shown in the subsequent images.
During this shrinkage, the droplet remains approximately spherical.
This is inconsistent with localized damage, as in that case, we would instead expect the droplet to close with a lenticular, crack-like shape.
Instead, damage appears to be distributed much as it would be in a ductile material.

\begin{figure}
\centering
\includegraphics[width=1\linewidth]{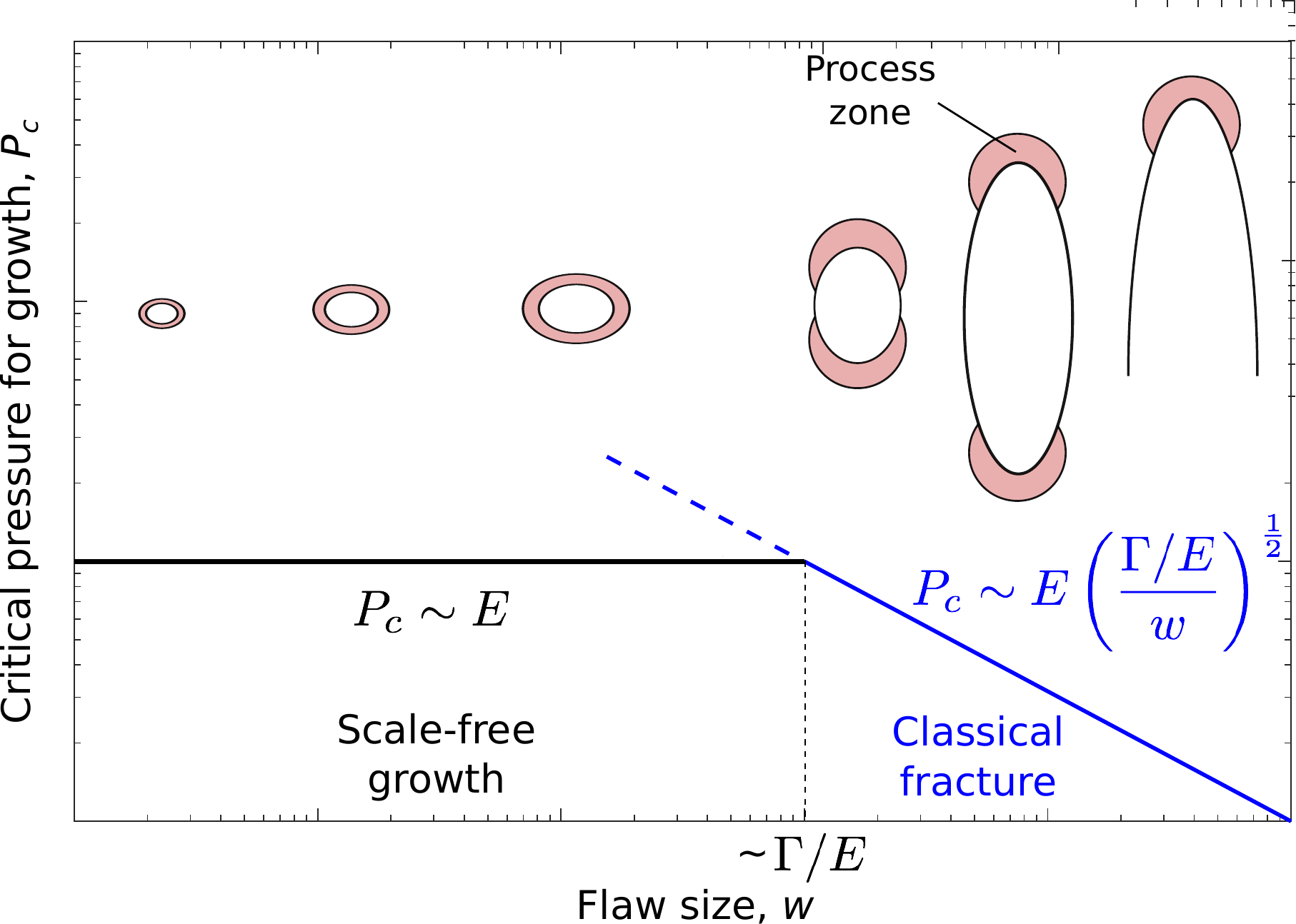}
\caption{A schematic of the suggested dependence of critical pressure for flaw growth, $P_c$ on flaw size, $w$. For small flaws (black), there is a scale-free regime with $P_c$ constant. Large flaws (blue) behave like classical cracks. We hypothesize that the cross-over between scale-free and crack-like regimes should occur when the process-zone size ($\sim \Gamma/E$ for large cracks) is comparable to the size of the flaw.
The schematic figures illustrate the anticipated change in cavity and process-zone morphology during growth. Note that we expect the flaw to change from aligned with, to perpendicular to the stretch direction when it reaches the crack-like regime.}
\label{fig:discussion}
\end{figure}

We can infer further information about how the damaged zone around droplets grows by comparing growth and shrinkage curves for different sized droplets.
Figure \ref{fig:itsfracture}d shows the typical evolution of the shape of various droplets that grow and shrink (due to slow diffusion out of the side of the sample) in silicone with $E=333$kPa and $\epsilon_x=60\%$ (c.f. Supplementary videos).
If we scale droplets' growth trajectory by their dimensions at their maximum size, $w_{\mathrm{max}}$ and $l_{\mathrm{max}}$, all of the data collapse onto a single trajectory, as shown in the inset.
This collapse shows that the whole growth process is self-similar -- and thus that the process zone must grow with the droplet, as shown schematically in Figure \ref{fig:discussion}.

\subsection{Cavity growth is independent of the fracture energy}

The experiments show several key features: i) cavities grow as smooth-walled spheroids, ii) they grow in a self-similar manner, and  iii) growth is accompanied by damage that is distributed around the cavity surface, rather than being localized to a crack tip. These are all at odds with crack-like growth.

We can rule out a dependence of cavity growth on the silicone's brittle fracture properties with a simple dimensional argument.
From above, $\alpha$ during initial growth is independent of the droplet's size.
Thus it only depends on $\epsilon_x$ and the silicone's material properties describing elasticity (e.g. $E$), fracture (the fracture energy, $\Gamma$) and damage (e.g. the stress at which inelasticity sets in, $\sigma_i$):
\begin{equation}
\alpha=f(\epsilon_x, \Gamma, E,\sigma_i...).
\label{eqn:a}
\end{equation}
$\Gamma$ has units of pressure $\times$ length, while the other material properties are either dimensionless, or have units of pressure.
Thus there is no dimensionally consistent way that $\alpha$ can depend on $\Gamma$, so growth must be independent of $\Gamma$.
This is intuitive as $\Gamma$ describes the fracture process where damage is localized to a crack tip -- which does not occur here.

\subsection*{Growth at constant pressure $\sim E$}

As well as showing that growth is independent of $\Gamma$, self-similar, non-brittle growth also shows that $P_c$ is constant during growth.
This matches previous theoretical results from the ductile void-growth literature, which showed that self-similar void growth occurs in elastic-plastic materials at constant driving stress (e.g. \cite{rice69,durb97,cohe10}).
This also explains the stability of droplets in our experiments.
If $P_c$ changes as droplets grow, then we would expect transport of fluorinated oil between droplets of different sizes -- i.e. ripening -- even when the temperature is held constant.
However in recent work, we observed no evidence of this \cite{roso19}.

This result could also explain why cavitation experiments in soft gels do not always give good agreement with the long-established result for elastic cavitation in incompressible neo-Hookean solids, that $P_c/E=5/6$ \cite{gent59,gent91,zimb07,kund09,hutc16}.
Our results show that growth pressure is constant, much like cavitation theory.
However, the exact value of $P_c$ will be determined by the consitutive relationship of the material at very large deformations, including inelastic and nonlinear elastic contributions.

\subsection*{Different regimes of cavity-growth behavior}

Our results suggest that there are different regimes of pressure required to open a flaw in a gel/elastomer, as shown in Figure \ref{fig:discussion}.
For small cavities, growth is scale-free, with $P_c\sim E$ constant.
For large cavities, growth is known to be crack-like, with damage localizing to a crack tip \cite{kund09,poul17}.
Then, linear-elastic fracture mechanics gives that $P_c\sim \sqrt{\Gamma E/w}$ (\emph{e.g.} \cite{lin04}).

We can naively predict the transition point between these two regimes by equating the two expressions for $P_c$ to find a crossover at $w\sim \Gamma/E$ (see Figure \ref{fig:discussion}).
This is consistent with our data.
In Table \ref{table} we report measured values of $\Gamma/E$.
These show that $w\ll\Gamma/E$ in all our experiments, so we expect self-similar growth.

The cross-over point, $\Gamma/E$, is interesting, as this \emph{elasto-adhesive length} is known to play an important role in soft fracture  \cite{cret16}.
In particular, it represents the effective process-zone size at the tip of a large crack \cite{hui03,cret16}.
Thus our results can be interpreted physically as showing that scale-free growth is expected at scales much smaller than this characteristic process-zone size.

This is completely analogous to metal cavitation.
Voids in ductile metals expand when $P_c\sim \sigma_y$, where $\sigma_y$ is the yield stress (c.f. Supplement, \cite{hill50}).
Large cracks will also fail by brittle fracture when $P_c\sim \sqrt{\Gamma E/w}$.
Equating these, we find a crossover when $w\sim \Gamma E/\sigma_y^2 \equiv L_p$. 
This is a well-established transition length-scale in ductile materials \cite{tayl08}, and also the size of the plastic process zone for large cracks \cite{dugd60}.

The main difference between the two types of material is in terms of scale.
$L_p$ in metals is typically macroscopic (e.g. $L_p\sim1\mathrm{cm}$ in steel \cite{cret16}).
Thus even macroscopic flaws in metals often grow in a ductile way, and void growth can be observed directly in experiments -- for example by post-examination of yielded samples.
Hence, metals are commonly considered as ductile materials.
However, in soft materials, $\Gamma/E$ is microscopic (see the Table).
Thus, these soft materials also exhibit non-brittle behavior, but this is much harder to observe as it takes place at much smaller scales.
By the time a cavity grows to a macroscopic size, it is in the brittle, crack-like regime where it will typically grow in a fast, unstable manner.
Hence, elastomers are typically considered as having a brittle failure response.

\subsection{Conclusions}

Every crack and cavity starts small.
Thus, understanding their nucleation and early growth is crucial to understanding how they develop.
Here, we have developed a new method that allows us to grow and shrink microscopic cavities in soft materials with precise volume control, revealing the key physics underlying cavity growth in soft materials.
We find that elastomers are not completely brittle materials, as commonly assumed.
Instead, small, growing cavities appear to have much more in common with void growth in ductile metals, being accompanied by distributed damage around the surface of the cavity, and growing at constant inflation pressure.
This rationalizes a number of experimental observations, including measurements of `flaw-insensitive' rupture in soft materials \cite{chen17}.
We hypothesize that this scale-free, inelastic behavior occurs in soft materials at scales smaller than the material length-scale $\Gamma/E$
-- provided surface tension effects are negligible \cite{styl17}.

The mechanism we describe opens up many interesting directions for future work, including fundamental questions about the behavior of small flaws in soft materials.
In particular, it will be important to develop new experimental and theoretical techniques to probe transitions between behavior at different length-scales. 
For example, we anticipate that one can extend cavitation techniques (e.g. \cite{zimb07,raay19}), to measure the critical cavitation pressure, $P_c$, as a function of cavity size during growth, and thus allow the testing of the hypothesis in Figure \ref{fig:discussion}.

Importantly, experiments like these could also be used to extract useful information about the large-strain behavior of materials.
Normally this is difficult to do with macroscopic experiments, as large samples break before they reach very high strains,
but our approach allows us to stably induce very large strains without fracture.
Thus it could be possible to use measurements of $P_c$, and the shapes of growing and shrinking droplets (like those in Figures \ref{fig:selfsimilar},\ref{fig:itsfracture}) to measure otherwise inaccessible material damage properties such as $\sigma_i$.
This will require a more detailed understanding of how damage occurs around a growing cavity, but this is seemingly an ideal topic for cutting-edge experimental techniques for imaging damage (\emph{e.g.} \cite{ducr14,cret17}), and numerical simulations including stress-softening or damage models.
Ultimately, a knowledge of how materials fail at high strains will give us insight into the structure-property relationships that determine how materials fail, paving the way to allowing us to design novel, tough materials (e.g. \cite{gong03,sun12,sun13}).

\subsection{Materials and Methods}
Our silicone gels consisted of a mixture of vinyl-terminated, silicone polymer chains (DMS-V31, Gelest) cross-linked with a methylhydrosiloxane-dimethylsiloxane copolymer (HMS-301, Gelest) with ratios of 69:1, 49:1 and 39:1 by mass \cite{styl15}.
Respectively, these had $E=71,333$ and $800$kPa.
Cross-linking was achieved by mixing in a small amount of Karstedt's catalyst (SIP6831.2, Gelest) -- approximately 0.01\% of the total mass of the sample.
Once mixed, degassed, and poured into moulds, samples were kept at 40$^\circ$C for 24 hours to ensure complete cross-linking.

We measured $E$ for the gels by indenting bulk samples (at least 10mm in depth) with a 1mm-radius, cylindrical indenter on a texture analyser with a 500g load cell (TA.XTPlus, Stable Microsystems).
We assume sample incompressibility (a good assumption for soft gels and elastomers \cite{styl14}), and then extract $E$ from the initial slope of the force-indentation curve (\emph{e.g.} \cite{styl15}).

We measured $\epsilon_x$ in stretched samples using placing marks on the samples. $\epsilon_x$ was then calculated by comparing the distance between marks during stretch, and after subsequent stretch release.

We measured $\Gamma$ using the Rivlin-Thomas pure-shear test \cite{rivl53,long16}.
100mm wide, 2mm-thick gel sheets were clamped between two long, straight clamps, with a distance of 20mm between the clamps.
We then extracted $\Gamma$ by comparing the  loading behavior of cracked, and crack-free  sheets, following \cite{sun12} (see also the Supplementary Information).

CARS microscopy was performed on a confocal Leica TCS SP8 microscope equipped with a tunable CARS laser (picoEmeraldS, APE Berlin), and a non-descanned external detector (Leica HyD, 600-725nm).
We used a 25x water-immersion objective (Leica HC FLUOTAR L 25x/0.95 W VISIR).
We visualized the silicone signal at a wavenumber of 2912$\mathrm{cm}^{-1}$, using a 1032nm Stokes beam, and a 793.8nm pump beam.

We thank Matteo Ciccotti, Anand Jagota and Edward Muir for helpful conversations, and Daniel King for advice on measuring fracture toughness. CARS microscopy experiments were performed using the ScopeM facilities at ETH Zurich. We thank Dr. Justine Kusch and Dr. Dorothea Pinotsi for help in performing these experiments. RWS is supported by the Swiss National Science foundation (Grant 200021-172827). JYK and BWM are supported by SKKU Global Challenge, Sungkyunkwan University, 2018. C.Y. Hui is supported by the National Science Foundation (Grant No. CMMI-1537087).

%

\end{document}